\documentstyle[psfig,sprocl]{article}

\def\be{\begin{equation}}
\def\ee{\end{equation}}
\def\bea{\begin{eqnarray}}
\def\eea{\end{eqnarray}}
\newcommand{\sla}{\!\!\!\!/ \,}

\begin{document}

\title{GLUON CONDENSATE AND PARTON PROPAGATION IN A QUARK-GLUON PLASMA}

\author{A. SCH\"AFER}

\address{Institut f\"ur Theoretische Physik, Universit\"at Regensburg,\\
93040 Regensburg, Germany\\E-mail: andreas1.schaefer@physik.uni-regensburg.de} 

\author{M.H. THOMA}

\address{Institut f\"ur Theoretische Physik, Universit\"at Giessen\\
35392 Giessen, Germany\\E-mail: markus.thoma@theo.physik.uni-giessen.de}


\maketitle\abstracts{A calculation of the thermal quark propagator 
is presented taking the gluon condensate above the critical 
temperature into account. The quark dispersion relation following
from this propagator is derived.}

\section{Introduction}

Finite temperature QCD can be used to describe properties of a
quark-gluon plasma (QGP) possibly created in the fireball of
relativistic heavy ion collisions and in the early Universe.
There are two different methods available, namely lattice QCD
and perturbation theory.

Lattice QCD is a non-perturbative method, which can be used
at all temperatures above and below the phase transition. However,
it is very difficult to calculate dynamical quantities or to consider 
a system at finite baryon density or in non-equilibrium.

Using perturbative QCD, on the other hand, one is able to compute
static as well as dynamical quantities at zero and finite baryon
density and even out of equilibrium. However, the strong coupling
constant $\alpha _s$ is small only for temperatures $T\gg T_c$.
Furthermore, infrared singularities show up in perturbative calculations
even within the hard thermal loop (HTL) resummation scheme
\cite{ref1}.

Therefore it would be desirable to construct non-perturbative
QCD Green functions. Here we suggest to include the gluon condensate
into the parton propagators. Condensates of gluons and quarks describe
the non-perturbative feature of the QCD ground state. They have been used
in QCD sum rules to describe hadron properties at zero and finite
temperature.  

Whereas the quark condensate is related to chiral symmetry breaking,
the gluon condensate
is associated with the breaking of the scale invariance. Therefore,
in contrast to the quark condensate, it does not vanish above the
phase transition. 


In the case of a pure gluon gas with energy density $\epsilon $ and 
pressure $p$ the gluon condensate can be related to the interaction 
measure $\Delta =(\epsilon -3p)/T^4$ via \cite{ref3}
\be
\langle G^2 \rangle _T=\langle G^2  \rangle _0-\Delta T^4,
\label{e5}
\ee
where $G^2\equiv (11\alpha _s/8\pi)\> : G^a_{\mu \nu}G_a^{\mu \nu}:$ 
and $\langle G^2  \rangle _0 = (2.5 \pm 1.0)\> T_c^4$.
Comparing $\langle G^2 \rangle _0$ with $\Delta T^4$, measured
on the lattice \cite{ref4} between $T_c$ and $4T_c$, one observes 
that the thermal gluon 
condensate $\langle G^2 \rangle _T$ above $T_c$ is different from 
zero and increases like $T^{\alpha}$ with $\alpha $ between 3 and 4. 

\section{Gluon Condensate and Quark Propagator}

As a first problem we want to study the influence of the gluon condensate 
on the quark and gluon propagation in the QGP \cite{ref4a}. At zero temperature
quark and gluon propagators containing condensates have been constructed
already$\>$\cite{ref5}. Here we will extend these calculations to finite
temperature, starting with the thermal quark propagator.

\begin{figure}[t]
\centerline{\psfig{figure=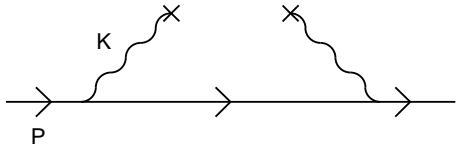,width=6cm}}
\caption{Quark self energy containing a gluon condensate.}
\end{figure} 

To lowest order in the gluon condensate the quark self energy is given
by the diagram of Fig.1. 
The most general ansatz for the quark self energy in the rest frame of 
the QGP is given by \cite{ref6}
\be
\Sigma (P)=-a(p_0,p)\> P\sla - b(p_0,p)\> \gamma _0,
\label{e6}
\ee
where we assumed a vanishing bare quark mass and used the notations
$P=(p_0,{\bf p})$, $p=|{\bf p}|$.

The scalar functions $a$ and $b$ can be expressed as
\bea
a & = & \frac{1}{4p^2}\> \left [tr\, (P\sla \Sigma ) - p_0\> tr\, 
(\gamma _0 \Sigma )\right ],\nonumber \\
b & = & \frac{1}{4p^2}\> \left [P^2\> tr\, (\gamma _0 \Sigma ) - p_0\> 
tr\, (P\sla \Sigma )\right ].
\label{e7} 
\eea

The most general ansatz for the non-perturbative gluon propagator
in Fig.1 at finite temperature is given by \cite{ref7}
\be
\tilde D_{\mu \nu} (K)=\tilde D_l(k_0,k)\> P_{\mu \nu}^l 
+ \tilde D_t(k_0,k)\> P_{\mu \nu}^t,
\label{e8}
\ee
where we have subtracted the bare gluon propagator as we are not
interested in perturbative corrections to the quark self energy. 
Consequently the gluon propagator (\ref{e8}) is gauge independent.
The projectors $P_{\mu \nu}^{l,t}$ project onto the longitudinal
and transverse degrees of freedom. 

Using the imaginary time formalism and expanding the quark propagator
in Fig.1 for small loop momenta, i.e. $k\ll p$ and $k_0=2\pi inT=0$,
we obtain
\bea
a & = & -\frac{4}{3}\, g^2\, \frac{T}{P^6}\int \! \frac{d^3k}{(2\pi )^3}\,
\left [\left (\frac{1}{3}p^2-\frac{5}{3}p_0^2\right )k^2\, \tilde 
D_l(0,k)+ \left (\frac{2}{5}p^2-2p_0^2\right )k^2\, 
\tilde D_t(0,k)\right ], \nonumber \\
b & = & -\frac{4}{3}\, g^2\, \frac{T}{P^6}\int \! \frac{d^3k}{(2\pi )^3}\,
\left [\frac{8}{3}p_0^2\, k^2\, \tilde D_l(0,k)+ \frac{16}{15}p^2\, k^2\, 
\tilde D_t(0,k)\right ].
\label{e9} 
\eea

The moments of the longitudinal and transverse gluon propagators
are related to the the chromoelectric and chromomagnetic condensates
via
\bea
\langle {\bf E}^2\rangle _T & = & \langle :G_{0i}^aG_{0i}^a:\rangle _T
= 8T\> \int \frac{d^3k}{(2\pi )^3}\> k^2\> \tilde D_l(0,k),\nonumber \\
\langle {\bf B}^2\rangle _T & = & \frac{1}{2}\> 
\langle :G_{ij}^aG_{ij}^a:\rangle _T
= -16T\> \int \frac{d^3k}{(2\pi )^3}\> k^2\> \tilde D_t(0,k).
\label{e10} 
\eea
These condensates can be extracted from the expectation values of the
space- and timelike plaquettes $\Delta _{\sigma ,\tau}$ computed
on the lattice \cite{ref4}, using
\bea
\frac{\alpha _s}{\pi }\> \langle {\bf E}^2 \rangle _T & = & \frac{4}{11}\>
\Delta _\tau\> T^4 - \frac{2}{11}\> \langle G^2\rangle _0,\nonumber \\ 
\frac{\alpha _s}{\pi }\> \langle {\bf B}^2 \rangle _T & = & -\frac{4}{11}\>
\Delta _\sigma\> T^4 + \frac{2}{11}\> \langle G^2\rangle _0.
\label{e11}
\eea

Combining (\ref{e6}), (\ref{e9}), (\ref{e10}), and (\ref{e11}) 
a gauge invariant expression for the quark self energy as a function of the 
measured plaquette expectation values and the zero temperature gluon condensate
is found. The effective quark propagator can be 
written by decomposing it according to its helicity eigenstates$\>$\cite{ref8}
\be
S(P)=\frac{1}{P\sla -\Sigma (P)}=\frac{\gamma _0-\hat p\cdot \vec \gamma}
{2D_+(P)} + \frac{\gamma _0+\hat p\cdot \vec \gamma} {2D_-(P)},
\label{e12}
\ee
where $D_\pm (P)=(-p_0\pm p)\> (1+a) - b$.

\section{Quark Dispersion Relations}

The dispersion relations of quarks and gluons are one of the most
important applications of thermal field theory \cite{ref6}. The quark
dispersion relations \cite{ref8} follows from $D_\pm (P)=0$. Using the lattice
results \cite{ref4} for the plaquette expectation values they have been 
determined numerically and are shown in Fig.2 for various temperatures. 

\begin{figure}[t]
\centerline{\psfig{figure=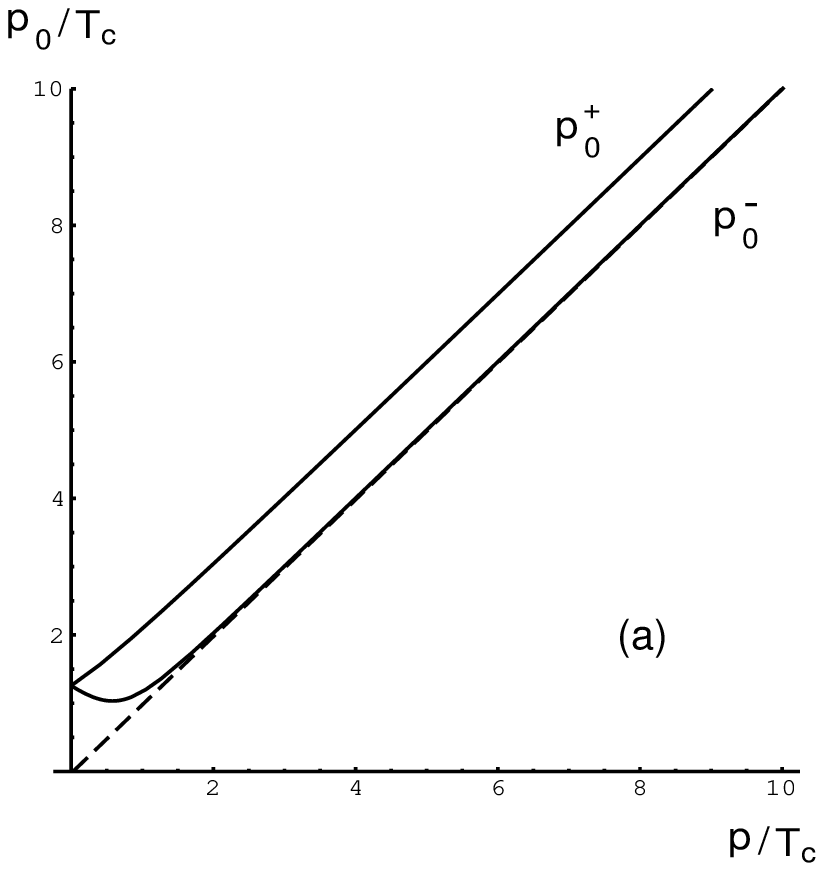,width=4cm}
\psfig{figure=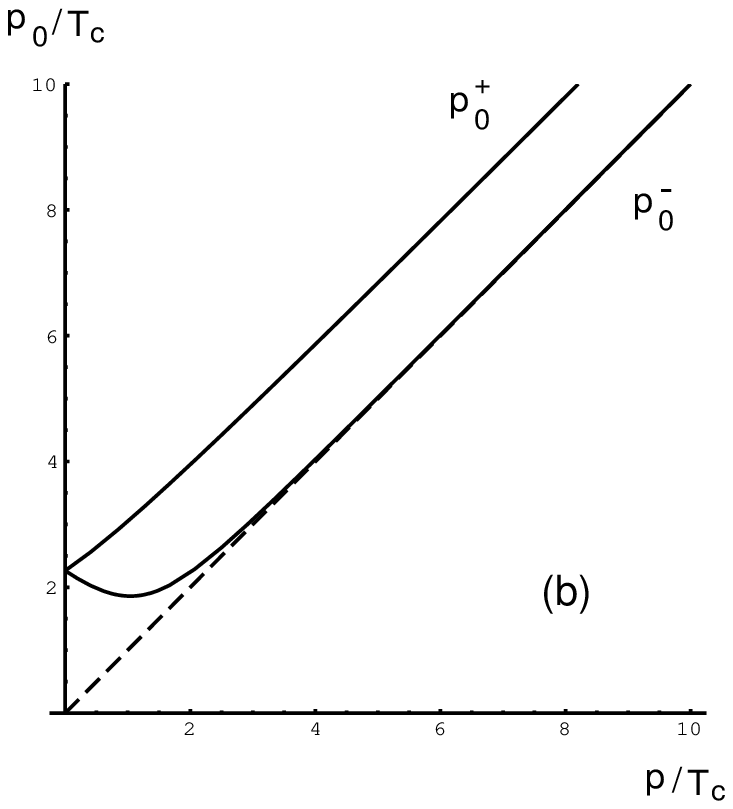,width=4cm}
\psfig{figure=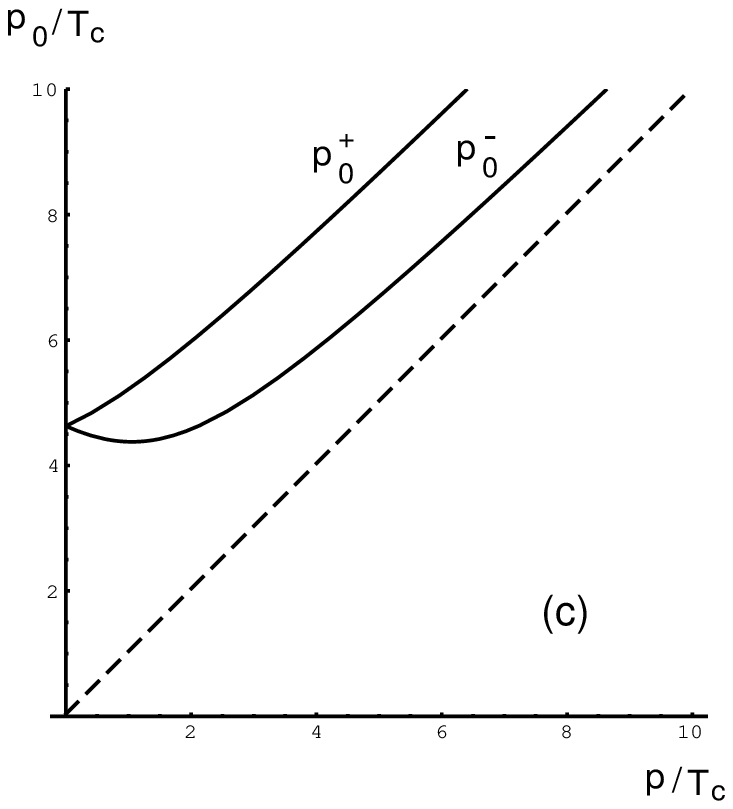,width=4cm}}
\caption{Quark dispersion relations at $T=1.1\, T_c$ (a), $T=2 \, T_c$ (b),
$T=4\, T_c$ (c) and dispersion relation of a non-interacting massless quark
(dashed lines).}
\end{figure}

The dispersions exhibit two massive quark modes. The upper branch comes 
from the solution of $D_+=0$ and the lower one from $D_-=0$. The lower
branch, showing a minimum, corresponds to a so-called plasmino, possessing 
a negative ratio of helicity to chirality, and is absent in the vacuum. 
The plasmino branch disappears for large momenta rapidly indicating
its collectivity.

At $p=0$ both modes start from a common effective quark mass, which is 
given by
\be
m_{eff}=\left [\frac{2\pi ^2}{3}\> \frac {\alpha _s}{\pi }\> \left (
\langle {\bf E}^2\rangle _T + \langle {\bf B}^2\rangle _T \right )
\right ]^{1/4}. 
\label{e13}
\ee
In the temperature range $1.1 T_c <T< 4T_c$ we found approximately
$m_{eff}=1.15 \> T$. 

The qualitative picture of this quark dispersion relation is very similar
to the one found perturbatively in the HTL limit \cite{ref8}. The main 
difference is the different effective mass, which is given by 
$m_{eff}=gT/\sqrt{6}$ in the HTL approximation.

%
%
%
 
\section{Conclusions and Outlook}

The gluon condensate above the critical temperature has been measured 
on the lattice. This condensate describes non-perturbative effects in 
the QGP. Following the zero temperature calculation we have constructed
an effective quark propagator containing the gluon condensate. Furthermore
we derived the quark dispersion relations from this effective propagator.
Similar as in perturbation theory (HTL approximation) we found two 
branches corresponding to a collective quark and a plasmino mode. Both 
branches start at the same effective quark mass. In contrast to the HTL 
approximation, where the effective quark mass is of order $gT$, we found
a mass proportional to $T$. 

As a possible application of this effective quark propagator we mention the
computation of the photon and dilepton production rates from the QGP.
For this purpose the photon self energy using effective quark propagators
should be evaluated. 

Finally we discuss briefly the possibility to construct a thermal gluon 
propagator for temperatures above $T_c$, containing the gluon condensate. 
This would be of interest, because one could study the screening behavior 
of such a non-perturbative propagator in the magnetic sector, where the 
absence of static magnetic sreening in the 
HTL resummed propagator leads to infrared singularities in perturbative 
calculations. 

As a consequence of the Slavnov-Taylor identities the gluon self energy 
including the gluon condensate should be transverse. Already at zero 
temperature this leads to the necessity to take ghost and higher order
condensates into account \cite{ref9}. Then one ends up with a complicated
expression for the gluon self energy containing unknown condensates.
Furthermore the gluon self energy turns out to be gauge dependent, which
leads to a gauge dependent gluon dispersion within this approximation,
rendering the physical meaning of this dispersion unclear.

\end{document}